\documentstyle[aps,prl,preprint]{revtex}
\newcommand{\gsim}{\lower.7ex\hbox{$\;\stackrel{\textstyle>}{\sim}\;$}}
\newcommand{\lsim}{\lower.7ex\hbox{$\;\stackrel{\textstyle<}{\sim}\;$}}

\def\tw{\theta_{\!\rm w\,}}
\def\sintw{\sin\tw}
\def\costw{\cos\tw}
\def\bm#1{{\mbox{\boldmath $#1$}}}
\def\tc{T_{\rm c}}

\begin{document}
\footnotesep=14pt
\begin{titlepage}
\begin{flushright}
\baselineskip=14pt
{\normalsize FERMILAB--Pub--97/257-A}\\
{\normalsize { hep-ph/9707513}}
\end{flushright}
\renewcommand{\thefootnote}{\fnsymbol{footnote}}
\baselineskip=24pt

\begin{center}
{\Large\bf
Electroweak Origin of \\ Cosmological Magnetic Fields}\\
\baselineskip=16pt
\vspace{0.75cm}

{\bf Ola T\"{o}rnkvist\footnote{\baselineskip=16pt
Present address: DAMTP, Univ.~of
Cambridge, Silver Street, Cambridge CB3~9EW, United Kingdom.
Electronic address: {\tt O.Tornkvist@damtp.cam.ac.uk}}}\\
\vspace{0.4cm}
{\em NASA/Fermilab Astrophysics Center},\\
{\em Fermi National Accelerator Laboratory},\\
{\em Batavia, Illinois~60510-0500, USA}
\vspace{0.3cm}
\vspace*{0.75cm}

July 21, 1997\\
Revised April 3, 1998\\
\end{center}
\baselineskip=20pt
\begin{quote}
\begin{center}
{\bf\large Abstract}
\end{center}
\vspace{0.2cm}
{\baselineskip=10pt Magnetic fields may have been generated in the electroweak
phase
transition through spontaneous symmetry breaking or through the subsequent
dynamical evolution of semiclassical field configurations.  Here I demonstrate
explicitly
how magnetic fields emerge spontaneously in the phase transition also when no
gradients of the Higgs field are present. Using a simple model,
I show that no magnetic fields are generated, at least initially,
from classical two-bubble collisions in a first-order phase transition.
An improved gauge-invariant definition of the
electromagnetic field is advocated which is more appropriate in the sense that
it never
allows electrically neutral fields to serve as sources for the electromagnetic
field. In
particular,  semiclassical configurations of the  $Z$ field alone do not
generate magnetic
fields. The possible  generation of magnetic fields in the decay of unstable
$Z$-strings
is discussed.}
\vspace*{8pt}
\noindent

\end{quote}
\end{titlepage}
\renewcommand{\thefootnote}{\arabic{footnote}}
\setcounter{footnote}{0}
\newpage
\baselineskip=24pt
\section{Introduction}

It is known that our galaxy and many other spiral galaxies
possess a
large-scale correlated magnetic field  with strength
of the order of $10^{-6}$ Gauss \cite{Kronberg}. In each case
the direction of the field seems to accord with the rotation axis of the
galaxy,
which suggests that it was generated by a dynamo mechanism in which
an initial field was amplified by the turbulent motion of matter
during the epoch of galaxy formation \cite{ZelPark}.
This mechanism usually requires a seed field of the order
of $10^{-21}$ Gauss or larger which is primordial in nature
(see, however, Ref.~\cite{Kulsrud} for an
alternative possibility). Various cosmological
explanations for a such a seed field have been suggested
\cite{Hogan,Baym,Sigl,VachaKep,Vachaspati,Grasso,Brandenburg,%
TurWid,misc}.
The present paper focuses on
scenarios in which a strong magnetic field of magnitude $10^{20}$-$10^{23}$
Gauss was
generated during the electroweak phase transition and was thenceforth diluted
by the
expansion of the universe to values appropriate for a seed field at the time
of onset of  galaxy formation.

There have been several models proposed in which the strong magnetic field
is produced by the turbulence of the conductive plasma
during the phase transition
\cite{Hogan,Baym}. In contrast, I shall restrict myself to
 mechanisms where the
magnetic field would be generated  directly from the dynamics of
the order parameter (the Higgs field)
and from
the gauge fields  in the process of breaking the electroweak symmetry
$SU(2)_{\rm L}\times U(1)_{Y}$ to $U(1)_{\rm EM}$. Such mechanisms include the
spontaneous
generation of magnetic fields, collisions of bubbles of  broken phase
in a first-order phase transition,
and the  formation and  dynamics  of non-topological defects.
In addition,
there are scenarios in which magnetic fields are produced by bound
pairs of  monopoles in standard and extended electroweak models
\cite{VachaKep},
but I shall not consider them here.

Vachaspati \cite{Vachaspati} has suggested  that strong magnetic fields may
emerge spontaneously in the phase transition
because the covariant derivatives of the Higgs field
in causally disconnected  regions must be uncorrelated. The electric
current that produces these fields can receive contributions from gradients of
the
phases of the Higgs field and charged vector-boson currents or both,
depending on which gauge is used.
Recently the electric current from the Higgs field was calculated in
 Ref.~\cite{Davidson}  and was found always
 to be zero.  For this reason, it was claimed that
no coherent magnetic fields are generated by the rolling Higgs field in the
electroweak
phase transition. I will show below that these statements are incorrect.

A useful tool in the investigation of magnetic phenomena and magnetogenesis
is the gauge-invariant definition of the electromagnetic field tensor
introduced in
Ref.~\cite{Vachaspati}. It has recently been employed
by Grasso and Riotto \cite{Grasso} in the study of semiclassical configurations
of the $Z$ and $W$ fields.
They discovered a set of puzzling paradoxes in which
the electrically neutral $Z$ field appears to act as a source for magnetic
fields.
In particular, it seemed that a magnetic field would always
be present along the internal axis of an electroweak $Z$-string.

These surprising and counter-intuitive results
have prompted me to reexamine the gauge-invariant
definition of the electromagnetic field tensor. I find that it is indeed
not suited to situations where the magnitude of the Higgs field deviates
from its vacuum value. I
propose a different definition of this tensor which,
in addition to
resolving the paradoxes,
proves to be a potent calculational tool. For example, it follows
immediately that
no magnetic field is generated initially from the classical dynamics of the
Higgs field in a collision between two bubbles in a
first-order electroweak phase transition.

The paper is organized as follows. In section II I describe the problems
with the
conventional gauge-invariant definition of the electromagnetic field
tensor and argue
why it should be modified. I then present an improved definition and
describe its general
properties. In section III I point out that the
contribution to
the electric current from the Higgs field actually does not vanish.
I go on to demonstrate that
in an arbitrary gauge one can always construct electrically charged field
directions in
the Lie algebra and corresponding charged vector-boson fields. The
current resulting from these fields is in general non-zero and
will give rise to electromagnetic fields.

In section IV I present an alternative description of
the spontaneous generation
of magnetic fields where the unitary
gauge is imposed. In this gauge there are no angular degrees of freedom
of the
Higgs field. Instead, the magnetic fields
arise from $SU(2)$ and $U(1)$ vector potentials that were present
already in the ground state of the symmetric phase.
As the $SU(2)_{\rm L}\times U(1)_{Y}$ symmetry breaks,
the vector potentials find themselves having
random non-vanishing components along
new physical directions of the Lie algebra
which are the eigenstates of mass and electric charge.
This reinterpretation confirms Vachaspati's original
proposal that magnetic fields can be generated spontaneously in the
electroweak phase transition \cite{Vachaspati}.

In section V, I show that no magnetic field is
generated initially from the classical dynamics of the Higgs field
in a collision between two bubbles in a
first-order electroweak phase transition.
This is shown for arbitrary difference
and relative orientation of the phases of the Higgs field in the
two
bubbles. The result is in stark contrast to that of the abelian $U(1)$
model, in which a field strength is present from the instant of
collision \cite{KibVil,Ahonen}.

In section VI the field configurations of the electroweak
$Z$-string \cite{Zref} and $W$-string \cite{embdef} are investigated,
using the redefined electromagnetic field tensor.
I verify that they carry neither magnetic fields nor electric currents.
In Ref.~\cite{Grasso} it was suggested that magnetic fields may be generated
in the decay of electroweak strings. In the case of the $Z$ string, the source
of
the magnetic field would be
charged $W$ fields which are initially present in the decay.
By constructing the unstable $W$ mode responsible for the decay,
I verify explicitly that a magnetic field is indeed generated.

Estimates of the strength and correlation length of the generated
magnetic
field are provided, for each mechanism separately, at
the end of sections IV, V and VI, respectively.

\section{Gauge-Invariant Definition of the Electromagnetic Field}

The conventional  gauge-invariant definition of the electromagnetic field
tensor in
the $SU(2)\times U(1)$ Yang-Mills-Higgs system is
given by \cite{Vachaspati}
\begin{eqnarray}\label{HooftSM}
{F^{{\rm em}}_{\mu\nu}} &\equiv& - \sintw{\hat \phi}^a F^{a}_{\mu\nu}  +
\costw F^{Y}_{\mu\nu}\nonumber\\
&-&
i\frac {\sintw}{g}\frac{2}{\Phi^{\dag}\Phi} \left[
\left({\cal D}_{\mu}{\Phi}\right)^{\dag}{\cal D}_{\nu}{\Phi} -
\left({\cal D}_{\nu}{\Phi}\right)^{\dag}{\cal D}_{\mu}{\Phi} \right],
\end{eqnarray}
where
$$
{\hat{\phi}}^{a} \equiv
\frac{\Phi^{\dag}\tau^{a}\Phi}{\Phi^{\dag}\Phi}~, \quad\quad
{\cal D}_\mu
= \partial_\mu - i\frac{g}{2} \tau^a W_\mu^a - i\frac{g'}{2}  Y_\mu
\equiv \partial_\mu - i \underline{A}_\mu~.
$$
This definition of $F^{{\rm em}}_{\mu\nu}$
has the attractive property that, in a ``unitary''
gauge
where $\Phi = (0,\rho)^\top$, ${\hat{\phi}}^{a}=-\delta^{a3}$,
 with $\rho$ real and positive, it reduces to the usual expression
$A_{\mu\nu}\equiv
\partial_\mu A_\nu - \partial_\nu A_\mu$  where $A_\mu =
\sintw W_\mu^3 + \costw Y_\mu$. This holds true, however,
only when the magnitude
 $\rho$ is a constant. For a general (positive) $\rho=\rho(x)$, it is easy to
show that
\begin{equation}\label{badf}
F^{{\rm em}}_{\mu\nu} = A_{\mu\nu} - 2 \tan\tw
(Z_\mu\partial_\nu \ln\rho - Z_\nu\partial_\mu \ln\rho)\quad\quad\mbox{(unitary
gauge)}
\end{equation}
with $Z_\mu =
\costw W_\mu^3 - \sintw Y_\mu$.

While such a definition certainly is possible, its physical consequences
become
highly disturbing when one considers the dynamical equation for
$F^{{\rm em}}_{\mu\nu}$ in this gauge, which
 takes the form \cite{Grasso}
\begin{eqnarray}\label{MaxwellUG}
\partial^{\mu} F^{\rm{em}}_{\mu\nu} &=&
- ie\left[ W^{\mu \dag}\left({\cal D}_\mu W_{\nu}-{\cal D}_\nu W_{\mu}\right)
-
\left({\cal D}_\mu W_\nu-{\cal D}_\nu W_\mu\right)^{\dag} W^{\mu}\right]
\nonumber \\
&-& ie \partial^\mu\left(W_\mu^{\dag} W_\nu - W_\nu^{\dag} W_\mu
\right)\nonumber\\
&-& 2\tan\tw ~\partial^\mu\left(Z_\mu\partial_\nu\ln\rho(x) -
Z_\nu\partial_\mu\ln\rho(x)\right)\quad\quad\mbox{(unitary gauge)}~.
\end{eqnarray}
Here $W_\mu^{\dag}$ and $W_\mu\equiv (W^1_\mu - i W^2_\mu)/\sqrt{2}$  are the
charged vector bosons, and ${\cal D}_\mu W_\nu \equiv (\partial_\mu - i g
W^3_\mu)
W_\nu$.

{}From Eq.~(\ref{MaxwellUG}) one would infer that
an electromagnetic field could be generated from currents involving the
fields
$Z_\nu$ and $\rho$. From most points of view such a result seems
absurd since, in the unitary gauge, $Z_\nu$ and $\rho$ are electrically
neutral. In fact, the charge operator $(\underline{\bf 1} + \tau^3)/2$
annihilates
$(0,\rho)^{\top}$ and commutes with the $Z$ direction in the Lie algebra,
$T_Z \propto \cos^2\tw \tau^3 - \sin^2\tw\underline{\bf 1}$. The
fields $Z_\nu$ and $\rho$ remain neutral also
when  $\rho$ is coordinate-dependent because
the form of the charge operator can depend only on the choice of gauge.
The change from
$\rho=$ constant to $\rho=\rho(x)$ does not constitute a change of gauge,
since
no angular degrees of freedom of
the Higgs field are involved.

The definition
(\ref{HooftSM}) thus implies that  electromagnetic fields can be produced by
neutral currents. A more reasonable and practical
definition should exclude this possibility.

Through a slight modification of a definition given by 't Hooft \cite{tHooft}
for the $SO(3)$ Georgi-Glashow model one obtains
an improved gauge-invariant definition of the electromagnetic field tensor,
\begin{equation}\label{goodone}
{\cal F}^{{\rm em}}_{\mu\nu} \equiv - \sintw{\hat \phi}^a F^{a}_{\mu\nu}  +
\costw F^{Y}_{\mu\nu} +
\frac {\sintw}{g}  \epsilon^{abc}
{\hat{\phi}}^{a} (D_\mu{\hat{\phi}})^{b} (D_\nu {\hat{\phi}})^{c}~,
\end{equation}
where $(D_\mu{\hat{\phi}})^{a} = \partial_\mu\hat{\phi}^a + g
\epsilon^{abc} W_\mu^b\hat{\phi}^c$. This definition depends on the Higgs
field
only through the
unit vector $\hat{\phi}^a$ which is independent of the magnitude $\rho =
(\Phi^{\dag}\Phi)^{1/2}$.
Therefore, the problematic
terms in eqs.~(\ref{badf}) and (\ref{MaxwellUG}) involving gradients
of  $\rho$ will not appear in the unitary gauge, where the field tensor now
always
reduces to
the familiar expression ${\cal F}^{{\rm em}}_{\mu\nu} = A_{\mu\nu}$.
An intricate interplay between the first and last term in eq.~(\ref{goodone})
ensures
that the electrically charged $SU(2)$ vector fields cancel (in any gauge),
leaving only the neutral component $- \sintw{\hat \phi}^a(
\partial_\mu W_\nu^a-\partial_\nu W_\mu^a)$. The definition (\ref{goodone})
has
been proposed earlier by Hindmarsh \cite{Hindmarsh},
but has not been applied before in the study of magnetic fields
in the electroweak phase transition.

It can be shown that the Bianchi identity
$\epsilon^{\mu\nu\alpha\beta}
\partial_\nu {\cal F}^{{\rm em}}_{\alpha\beta} = 0$ is satisfied everywhere
except
along world lines around which $\hat{\phi}^a$ takes ``hedgehog'' configurations
\cite{tHooft}. This ensures that there is no magnetic charge or
magnetic current in the absence of magnetic monopoles. The
conventional definition, eq.~(\ref{HooftSM}), does not have this property.

Repeating the calculation done in Ref.~\cite{Grasso} for the
 field tensor of eq.~(\ref{HooftSM}),
one may derive the field equation for
the redefined field tensor ${\cal F}^{{\rm em}}_{\mu\nu}$
using the equations of motion for $F^a_{\mu\nu}$
and $F^Y_{\mu\nu}$ and a few Fierz identities. One thus obtains
\begin{equation}\label{Fdyn}
\partial^\mu {\cal F}^{{\rm em}}_{\mu\nu} =  j^{\rm e}_\nu \equiv - \sintw
(D^{\mu}{\hat \phi})^a  F^{a}_{\mu\nu}  +
\frac{\sintw}{g} \partial^\mu\left[ \epsilon^{abc}
{\hat{\phi}}^{a} (D_\mu{\hat{\phi}})^{b} (D_\nu
{\hat{\phi}})^{c}\right] ~,
\end{equation}
where $j^{\rm e}_\nu$ is the gauge-invariant electric current.

It should be remarked that no physics is affected by using one definition
of the electromagnetic field rather than the other. In fact, in any
chosen gauge the field configuration is completely specified
by the components $W_\mu^a$ and $Y_\mu$ of the gauge potentials,
as defined by their occurrence in the covariant derivative ${\cal
D}_\mu$, together with the 4 real components of the Higgs field.
Two observers, using different definitions (\ref{HooftSM}) and
(\ref{goodone}) of the electromagnetic
field,
may then disagree on whether this same field configuration constitutes an
electromagnetic field or not. Clearly, this does not affect the
subsequent evolution of the field configuration. In the absence of
topological defects, it will evolve into
a state with uniform magnitude of the Higgs field, where the two
definitions coincide.

The choice of definition
is,
however, important for the interpretation, description and
understanding of physical processes
whenever $\Phi^{\dag}\Phi$ is not constant. In particular, one
should be aware that it may be meaningless to make strong claims about
the presence or absence of magnetic fields in situations that
involve a non-uniform magnitude of the Higgs field, unless one
is careful to specify which definition of the electromagnetic field
tensor is used.

In this paper, I adopt the modified definition (\ref{goodone}) which ensures
that there is no magnetic charge or magnetic current and that no
electromagnetic
field is generated from electrically neutral sources.
Even so, one should
remember
that  there is no exact standard by which definition (\ref{HooftSM}) would be
incorrect.

In Ref.~\cite{Grasso} it was stated that, because of the last term of
eq.~(\ref{MaxwellUG}),
 the formation of a magnetic field is
always associated to the appearance of a semiclassical $Z$-configuration.
As is seen from the above arguments, such a statement depends on the definition
of
the electromagnetic field.   In the view of the modified definition,
eq.~(\ref{goodone}),
no magnetic field would accompany the neutral-charge configuration.

\section{A Non-Vanishing Electric Current}

It was originally suggested by Vachaspati \cite{Vachaspati} that
electromagnetic
fields may emerge in the electroweak phase transition  through the
process of spontaneous symmetry breaking. The principal idea is that
as the Higgs field magnitude $\rho=(\Phi^{\dag}\Phi)^{1/2}$ becomes non-zero
in
the phase transition, the covariant derivative ${\cal D}_\mu\Phi\equiv
(\partial_\mu - i \underline{A}_\mu)\Phi$ cannot remain
 everywhere zero, because that  would imply an inexplicable
correlation of phases and gauge fields over distances greater than the
causal horizon distance
at the time of the phase transition.

In the much simpler case of a  global $U(1)$ symmetry
(i.e.~with  the gauge potential $A_\mu$ set to zero), an instructive analogy
can be made with the phase transition in superfluid He$^4$ \cite{Zurek}.  When
such a
system is rapidly quenched, the complex field $\Phi$
emerges from the false $\Phi\equiv 0$ ground state
 attempting to find a new true minimum
on the circle $|\Phi|=v$ but is
forced to assign values for its phase more quickly than the time it takes
information to
propagate across the container (given by the speed of ``second sound'').
Gradients of the phase thus appear and, because
the fluid velocity is proportional to the gradient of the phase, a flow is
generated.

The analogy with the superfluid has sometimes led to the misinterpretation
that
magnetic fields in the electroweak phase transition are generated
only by gradients of the
phases of the Higgs field.
Recently, it was claimed \cite{Davidson} that the electric current
resulting from Higgs gradients is always
zero, and that for this reason  no magnetic field would be
produced
during the phase transition due to spontaneous symmetry breaking.
As I will explain below, these conclusions were contingent
upon using an incomplete expression for the electric current
from the Higgs field as well as neglecting the
electric current from charged vector bosons\footnote{The latter
point was also made in Ref.~\cite{Grasso}.}.  In general
the electric current receives contributions both from
charged vector fields and from gradients of the
phases of the Higgs field.
For example, in section \ref{spont} it is shown that
magnetic fields  emerge spontaneously in the electroweak
phase transition also when no gradients of the Higgs
field are present.

Let us begin by considering the gauge-{\em co}variant
charge operator proposed in
Ref.~\cite{Davidson},
\begin{equation}\label{Qdef}
Q=-\frac{1}{2} \hat{\phi}^a \tau^a + \frac{Y}{2}~, \quad
\hat{\phi}^a =  \frac{\Phi^{\dag}\tau^a\Phi}{\Phi^{\dag}\Phi}~,
\end{equation}
where I define the hypercharge $Y$ of the Higgs doublet to be  $+1$.
This operator
has the property that $Q\Phi = 0$,
which can be understood as follows:
Due to gauge freedom, one
may represent the Higgs field of the vacuum state in any ``coordinate system''
of
choice through applying a gauge transformation to  $(0,v)^{\top}$.
This would not constitute an active, physical change of the state, but merely a
change of basis of the Lie algebra and its representations.
In the unitary gauge the vacuum state is represented by
\begin{equation}\label{univacuum}
\Phi_0=\left(\begin{array}{c} 0\\ v \end{array}\right)~,\quad
\underline{A}_\mu =
 \partial_\mu\lambda(x)\,Q~,\quad
\lambda\in{\bf R}~,\quad Q=Q_0\equiv\frac{1}{2}(\underline{\bf 1} + \tau^3)~,
\end{equation}
where the $u(1)$ ``pure-gauge'' form of  $\underline{A}_\mu$ is the most
general
expression for which ${\cal D}_\mu\Phi_0$ and
 the field tensors $F^a_{\mu\nu}$ and $F^Y_{\mu\nu}$
vanish. This vacuum state can be equivalently re\-expressed as
\begin{equation}\label{arbvacuum}
\Phi_0= \frac{\Phi~}{(\Phi^{\dag}\Phi)^{\frac{1}{2}}}\, v,~
\Phi(x)~{\rm arbitrary},~\quad
\underline{A}_\mu = \partial_\mu\lambda(x)\,Q
- i(\partial_\mu V)  V^{\dag} ~
\end{equation}
through a gauge transformation $\Phi_0\to V\Phi_0$ with $V\in SU(2)$ defined
by
\begin{equation}\label{Vdef}
V=\frac{1}{ (\Phi^{\dag}\Phi)^{\frac{1}{2}}} \left(
\begin{array}{cc}
(i\tau^2\Phi)^{\ast}&~ \Phi \end{array}\right)~.
\end{equation}
Under this transformation, $Q_0\to Q\equiv VQ_0 V^{\dag}$.
It can be checked that
this definition of $Q$ agrees with eq.~(\ref{Qdef}).

We see that when the charge operator $Q$ is defined covariantly
as in eq.~(\ref{Qdef}),
$\Phi$ is always proportional
to  the vacuum Higgs field $\Phi_0$ with a real factor.
Thus, $\Phi$ in this formulation
is always electrically neutral. The end result is a reformulation of
the unitary gauge in an arbitrary basis.

Let us now focus on the issue of the electric current.
In Ref.~\cite{Davidson} a current involving the Higgs field was
derived using the relation $j^\nu \equiv -\partial {\cal L}_H /\partial
A_\nu$, where ${\cal L}_H = ({\cal D}_\mu \Phi)^{\dag}
{\cal D}^\mu \Phi$ is the Higgs kinetic term in the Lagrangian and
$A_\nu$ is the Lie-algebra component along $Q$.
The resulting expression,
\begin{equation}\label{jHiggs}
j^\nu = i e [ \Phi^{\dag} Q {\cal D}^\nu \Phi - ({\cal D}^\nu \Phi)^{\dag}
Q\Phi]~,
\end{equation}
is zero by virtue of $Q\Phi =0$. This current, however, is
not the electric current, because
the $A_\nu$ used here is not the vector potential
whose curl gives ${\cal F}^{\rm em}_{\mu\nu}$.
A vector potential
with such a property can in fact be constructed \cite{prep}.
The electric current corresponding to ${\cal F}^{\rm em}_{\mu\nu}$
is given by eq.~(\ref{Fdyn}) and includes
a contribution from gradients of the phases of the Higgs field.
 Therefore, magnetic fields can
be produced by the classically evolving Higgs field in the electroweak
phase transition.

Magnetic fields also arise from charged
vector-boson currents
in the absence of gradients of the Higgs field.
In the remainder of this section I shall
construct the charged vector fields for an arbitrary choice
of  $\Phi$ in eq.~(\ref{arbvacuum})
and proceed to show that they give rise to an electric current which
in general is non-zero.

The charged vector-boson fields can be found by
determining the $SU(2)\times U(1)$ Lie-algebra
 eigenstates under the adjoint action of the gauge-covariant charge operator
$Q$. After some algebra and using a series of Fierz identities,
one can readily verify that
\begin{eqnarray}\label{comms}
&[Q, T_{\pm}] = \pm T_\pm~, \quad [Q,T_3]=[Q,\underline{\bf 1}]=0~, & \\
&[T_3, T_{\pm}] = \pm T_\pm~,  \quad [T_+,T_-]=2 T_3~,&
\end{eqnarray}
where
\begin{equation}\label{Tdef}
T_+\equiv \frac{(-i\Phi^{\dag} \tau^2)^{\top}
\Phi^{\dag}}{\Phi^{\dag}\Phi}~,\quad
T_-\equiv \frac{ \Phi (i\tau^2\Phi )^{\top}}{\Phi^{\dag}\Phi}=T_+^{\
\dag}~,\quad
T_3 \equiv -\frac{1}{2} \frac{\Phi^{\dag}\tau^a\Phi}{\Phi^{\dag}\Phi}\tau^a~,
\end{equation}
and $Q = T_3 + Y/2$.
Thus $T_+$ and $T_-$ are the generators of the Lie algebra
corresponding to charged field directions.  Using $T_\pm=T_1\pm i T_2$ we can
write
\begin{equation}\label{aingen}
\underline{A}_\mu = g
\tilde{W}^{a}_\mu T_a+ \frac{g'}{2} \tilde{Y}_\mu \underline{\bf 1}
+\partial_\mu\lambda(x)\,Q
- i(\partial_\mu V)  V^{\dag}~,
\end{equation}
where $\tilde{W}^{a}_\mu=\tilde{Y}_\mu = 0$ corresponds to the vacuum,
eq.~(\ref{arbvacuum}).
Under an SU(2) gauge transformation $\Phi\to U\Phi$
the generators $T_a$, $a=1,2,3, (+,-)$, transform according to the adjoint
representation
 $T_a\to UT_aU^{\dag}$,  and it can be shown that the fields $\tilde{W}^a_\mu$
are
gauge invariant. Furthermore, the field
tensor components $\tilde{F}^a_{\mu\nu} = 2{\rm Tr} (T_a
\underline{F}_{\mu\nu})/g$
are invariant
under general  $SU(2)\times U(1)$ gauge transformations.

The important point is that, in general, there will be charged vector-boson
fields $\tilde{W}_\mu\equiv (\tilde{W}^1_\mu-i\tilde{W}^2_\mu)/\sqrt{2}$
 and $\tilde{W}^{\,\dag}_\mu$ present
regardless of what gauge we choose for the vacuum, corresponding to the
components of the Lie algebra along $T_+$ and $T_-$.
I shall now show that these charged fields give rise to an electric
current and therefore magnetic fields.
First, let us evaluate the electromagnetic field tensor. Inserting the
components of $\underline{A}_\mu$ and
$\underline{F}_{\mu\nu} = \partial_\mu \underline{A}_\nu -
\partial_\nu \underline{A}_\mu  - i [\underline{A}_\mu,\underline{A}_\nu ]$
into eq.~(\ref{goodone}),
one finds after rather lengthy calculations
that the derivatives $\partial_\mu T_{a}$ in the first term cancel against
the last term, and we retrieve
\begin{equation}\label{goodFmunu}
{\cal F}^{{\rm em}}_{\mu\nu} = \sintw
 (\partial_\mu \tilde{W}^{\,3}_\nu - \partial_\nu \tilde{W}^{\,3}_\mu) +
\costw F^Y_{\mu\nu}~.
\end{equation}
Turning next to the field equation for ${\cal F}^{{\rm em}}_{\mu\nu}$,
eq.~(\ref{Fdyn}), insertion and yet more algebra produces
\begin{eqnarray}\label{Maxwell}
\partial^{\mu} {\cal F}^{\rm{em}}_{\mu\nu} &=&
- ie\left[ W^{\mu \dag}\left({\cal D}_\mu
W_{\nu}-{\cal D}_\nu
W_{\mu}\right) -
\left({\cal D}_\mu {W}_\nu-{\cal D}_\nu
{W}_\mu\right)^{\dag}
{W}^{\mu}\right]  \nonumber \\
&-& ie \partial^\mu\left({W}_\mu^{\dag} {W}_\nu -
{W}_\nu^{\dag} {W}_\mu
 \right)~,
\end{eqnarray}
where, as a final step, the ``tilde'' (\~{}) accents were
omitted.
This is exactly the expression (\ref{MaxwellUG}) obtained in the unitary
gauge,
but without the objectionable last term, as was discussed in the previous
section.

I have thus established that the treatment of Ref.~\cite{Davidson} is
equivalent to a treatment in the unitary gauge, where the Higgs field
possesses
no angular degrees of freedom. These degrees of freedom are absorbed into  the
vector bosons. However, the current from charged vector
bosons was omitted in \cite{Davidson}. In general this current, given by
eq.~(\ref{Maxwell}), is
non-zero and will give rise to electromagnetic fields.
In the next section we shall see an example of how this can happen.

\section{Spontaneous Generation of Magnetic Fields}
\label{spont}
Previous descriptions \cite{Vachaspati,Davidson}
of the spontaneous generation of magnetic fields in
the electroweak phase transition have
borrowed from the analogy with
superfluids in that they attribute the magnetic fields to the presence of
gradients of phases of the Higgs field. I present here an alternative
description of magnetogenesis where the unitary gauge is imposed.
In this gauge, there are no angular degrees of freedom of the Higgs field.
Instead, the magnetic fields arise from $SU(2)$ and $U(1)$ vector potentials
that were already present in the ground state of the
symmetric phase. As the
$SU(2)_{\rm L}\times U(1)_{Y}$ symmetry breaks, these
vector potentials find
themselves having random non-vanishing components along new
physical directions of the Lie algebra which are the
eigenstates of mass and electric charge.

When the symmetry breaks, unstable non-topological defects such as W-strings
and Z-strings  typically form carrying large fluxes of gauge fields.
In the
core of these defects the Higgs field $\Phi$ goes to zero, at which points the
unitary gauge is ill-defined. For now, I shall
consider a region of space where such defects are absent. Non-topological
defects will be considered in more detail in section \ref{topdef}.

In the symmetric phase, the vacuum state of electroweak model is characterized
by  $\Phi\equiv 0$, $F^a_{\mu\nu} = F^Y_{\mu\nu} = 0$. Surely,
 in the
high-temperature electroweak plasma there will
be fluctuations around the vacuum values, but these fluctuations are expected
to have
a small correlation length of the order $(2\pi T)^{-1}$, and we are primarily
interested in a mechanism that
may generate magnetic fields correlated on a larger scale. The macroscopic
spatial average of
$F^a_{\mu\nu}$ and $F^Y_{\mu\nu}$ on such a scale will also vanish, and
therefore
 the  Lie-algebra
valued vector potential is a Maurer-Cartan form
\begin{equation}
\label{symmvac}
\underline{A}_\mu = - i(\partial_\mu \Omega)  \Omega^{\dag} ~ \equiv
 - i(\partial_\mu U(x))  U(x)^{\dag} +\partial_\mu \chi(x)\,\underline{\bf
1}~,
\end{equation}
where $\Omega\in SU(2)\times U(1)$, $U\in SU(2)$ and $\chi\in{\bf R}$.
The group-valued function $\Omega$ maps to
the group manifold $S^3\times S^1$,
the direct product of a three-sphere and a circle, and
is completely arbitrary. Because the energy is independent of the space
dependence
of $\Omega(x)$, there is no reason that $\Omega$ should be uniform over
space.

Let us now consider the process of symmetry breaking, and for simplicity use
the unitary gauge in the broken state.
We shall assume that the temporal component $\underline{A}_0$
is a continuous function of the space coordinates
and require in addition
that the ``electric fields'' $F^a_{0i}$ and $F^Y_{0i}$
be everywhere finite.
Then the
spatial components $\underline{A}_i$, $i=1,2,3$, are continuous
functions of time, and
the initial $\underline{A}_i$
immediately after the phase transition are  given
by eq.~(\ref{symmvac}).
In general,
$\underline{A}_i$
will not be aligned with the vector potential of the broken-symmetry vacuum,
eq.~(\ref{univacuum}). This would happen only in the special case when
$\Omega$
is restricted to the embedded circle
$\Omega = e^{i\lambda(x)Q}$ as $x$ covers space.
For other choices of $\Omega$, for example
$\Omega = e^{i\xi(x) \tau^1}$, it is easy to check that
there will be physical,
 electrically charged $W$-boson fields present immediately after the phase
transition.
The ensuing state is a coherent
semi-classical field configuration which cannot be constructed from the
new  vacuum by perturbative means.

Let us now look at a concrete example of how the magnetic field is
generated.
The condition
\begin{equation}
\label{f3zero}
0 = F^3_{kl}\equiv  (\partial_k W^3_l - \partial_l W^3_k)
 + g \,(W^1_k W^2_l - W^1_l W^2_k)
\end{equation}
can be satisfied if both the first term and the second term are non-zero but
cancel exactly. The first term  enters  in the
unitary-gauge  definition of
the magnetic field in the broken phase:
\begin{equation}
\label{bdef}
A_{kl}= \sin\theta_w\, ( \partial_k W^3_l - \partial_l W^3_k)
+ \cos\theta_w  F^Y_{kl}~,
\end{equation}
where in our case $F^Y_{kl}=0$.

The emerging magnetic field can therefore be traced to  a  ``random''
partitioning of fields into the two cancelling terms of eq.~(\ref{f3zero}).
In the symmetric phase, these terms had no independent physical meaning,
and fields could be moved from one to the other through arbitrary gauge
transformations while keeping $F^3_{kl}$ zero.  When the symmetry
is broken, the terms take on a new physical meaning. The first term in
(\ref{f3zero}) has  components along $A_{kl}$ as well as along
$Z_{kl}=\partial_k Z_l - \partial_l Z_k$. The second term
in (\ref{f3zero}) can be written
\begin{equation}
\label{wpairs}
ig(W_k^{\dag} W_l - W_l^{\dag} W_k)
\end{equation}
in terms of the charged $W$ fields.
It is now apparent that there can be no spontaneous
generation of magnetic fields in the electroweak phase transition without
the simultaneous generation of charged W-boson currents which act as
the only source (in the unitary gauge)
for that magnetic field. In fact, the field equation for
the electromagnetic field in the unitary gauge, when $F^a_{\mu\nu}=0$,
is\footnote{\baselineskip=14pt It should be noted that
 when the two terms in equation
 (\ref{f3zero}) are non-zero, a state with $F^a_{\mu\nu}=0$ does not
remain an exact solution
in the broken phase because of the mass terms that appear there.}
\begin{equation}
\label{bsource}
\partial^\mu A_{\mu\nu} = -i e \partial^\mu
(W_\mu^{\dag} W_\nu - W_\nu^{\dag} W_\mu)~.
\end{equation}
The term on the right-hand side of this equation is the
magnetization current corresponding to the anomalous magnetic dipole moment
of the $W$ boson \cite{Ambjorn,MacDowell,thesis}.
The initial magnetic field can therefore be viewed as being entirely comprised
of magnetization of the vacuum due to $W$ bosons. This state has previously
been investigated in the context of the QCD vacuum \cite{Ambjorn2}.

Let us now see explicitly
how the two terms of  eq.~(\ref{f3zero}) obtain non-zero values
from a random vector potential in the symmetric ground state. Because
$F^a_{\mu\nu}= F^Y_{\mu\nu}=0$ the initial gauge potential must be given by
eq.~(\ref{symmvac}). The most general $SU(2)\times U(1)$-valued function
 $\Omega$
can be written
\begin{equation}
\label{uopar}
\Omega(x)= e^{i\lambda/2}\!\left({\begin{array}{ll}
e^{i (\lambda/2-\beta)} \cos\omega &~~ -e^{i(\alpha-\lambda/2)} \sin\omega\\*
e^{i(\lambda/2-\alpha)} \sin\omega & ~~e^{i (\beta-\lambda/2)} \cos\omega
\end{array}}\right)
= e^{i\lambda/2} U~,\quad U\in SU(2)~.
\end{equation}
The $su(2)$ algebra part of the gauge potential is given by $W^a_\nu \tau^a =
-(2 i/g) (\partial_\nu U) U^{\dag}$. The curl of its components
 can be calculated from
\begin{equation}
\label{trick}
\partial_{[\mu} W^a_{\nu]}\tau^a = {{2i}\over{g}}\left( \partial_{[\mu}
U\right)
\left(\partial_{\nu]}U^{\dag}\right)~,
\end{equation}
where $[\mu\ldots\nu]$ indicates antisymmetrization, using the trace
identity
${\rm Tr\,}\tau^a\tau^b = 2 \delta^{ab}$.
One then finds
\begin{equation}
\label{result}
\partial_{k} W^3_{l} - \partial_{l} W^3_{k}=
{{2}\over{g}} \sin 2\omega
\left( \omega_{[,k} \alpha_{,l]} + \omega_{[,k} \beta_{,l]}
-\omega_{[,k} \lambda_{,l]}\right)
= -g (W^1 _k W^2_l - W^1_l W^2_k)~,
\end{equation}
where a comma denotes partial differentiation.
Here we see that the two terms of eq.~(\ref{f3zero}) have the opposite
sign
and in general assume non-zero values that vary as one changes the
group-valued function $\Omega$. One of these terms gives rise
to the magnetic field, according to
eq.~(\ref{bdef}):
\begin{equation}
\label{magspont}
A_{kl} =
{{2\sintw}\over{g}} \sin 2\omega
\left( \omega_{[,k} \alpha_{,l]} + \omega_{[,k} \beta_{,l]}
-\omega_{[,k} \lambda_{,l]}\right)~.
\end{equation}
Thus, in this semiclassical description
it is a random vector potential
in the symmetric phase that gives rise
to a magnetic field in the broken phase.  In this sense, the magnetic field
was already present in the ground state of the
symmetric phase, but took on a different physical meaning
after the symmetry was broken and
eigenstates of mass and electric charge became well-defined.

We must now address the issue of gauge invariance. So far
we have used the unitary gauge to calculate the magnetic field
resulting from a
vector potential $\underline{A}_\mu$ expressed as the Maurer-Cartan form
(\ref{symmvac}). This potential is, however,
gauge-dependent, so it is necessary to show that the magnetic field
generated is independent of our choice of gauge.

Let us therefore pick an arbitrary gauge in which the
vector potential in the symmetric phase
is some particular function $\underline{A}_k=-i(\partial_k\Lambda)
\Lambda^{\dag}$ with $\Lambda\in SU(2)\times U(1)$.
Then $\Lambda$ is uniquely determined up to right-multiplication
($\Lambda\to\Lambda M$) by an arbitrary constant group element $M$.
By continuity,
the vector
potential is the same in the broken phase immediately after the phase
transition. In addition, we obtain in the broken phase
some isospin orientation of the
Higgs field, which in the same gauge
can be characterized by the matrix $V\in SU(2)$
defined by eq.~(\ref{Vdef}), such that $\Phi=\mbox{$V\cdot
(0,\rho)^{\top}$}$. Now we can evaluate the magnetic field, either
directly
from the gauge-invariant definition (\ref{goodone}), or
equivalently by making a gauge
transformation to the unitary gauge using
the $SU(2)$ element ${V}^{\dag}={V}^{-1}$. In this gauge we obtain
the vector
potential $\underline{A}'_k = -i [\partial_k(V^{\dag}\Lambda)]
(V^{\dag}\Lambda)^{\dag}\equiv
-i(\partial_k \Omega)\Omega^{\dag}$, where $\Omega =
V^{\dag}\Lambda$. Let us now show
that the magnetic field resulting
from $\underline{A}'_k$ is gauge invariant.

Under a general gauge transformation $g$
given by $g=e^{i\xi/2} U$ with $U\in SU(2)$,
we have that $\Lambda\to g\Lambda$, while
$V\in SU(2)$ transforms as $V\to gVh^{\dag}$, where
\mbox{$h\equiv e^{i\xi Q}$}
is an electromagnetic $U(1)$ gauge
transformation. Here $Q=(1+\tau^3)/2$.
 Therefore, under the full gauge transformation, $\Omega\to
h\Omega$. The electromagnetic part of the vector potential then
changes only by a pure gradient, corresponding to the remaining gauge
symmetry of the broken phase.
Furthermore, $\underline{A}'_k$ is invariant
under the transformation $\Lambda\to\Lambda M$ for constant $M$,
so this ambiguity in the definition of $\Lambda$ has no significance.
  The resulting magnetic field, obtained
from eqs.~(\ref{uopar}) and (\ref{magspont}) with $\Omega =
V^{\dag}\Lambda$, is
therefore independent of which gauge was used originally to express
$\underline{A}_k$ and the Higgs field.

Given a gauge potential $\underline{A}_\mu$ of the form
(\ref{symmvac}) in a unitary gauge, with $\Omega$
given by eq.~(\ref{uopar}),
we may conversely set the gauge potential to zero by means of
a gauge transformation with group element $\Omega^{-1}$.
The phases will then reappear in the Higgs field, which becomes
\begin{equation}\label{Higgsphase}
\Phi = \rho\left(\begin{array}{c} e^{i\alpha} \sin\omega\\
e^{-i\beta} \cos\omega\end{array}\right)~.
\end{equation}
The phase $\lambda$ does not appear here because
the broken vacuum still has the electromagnetic $U(1)$ symmetry.

Therefore, as long as $F^a_{\mu\nu}$ can be considered to vanish,
one can give two {\em equivalent} descriptions of magnetogenesis in two
different gauges. (a) In a gauge where all vector potentials are
identically zero, the magnetic field arises spontaneously
from  the angular degrees of freedom of the Higgs field
and is given by the last term of eq.~(\ref{goodone}).
(b) In the unitary gauge, with $\Phi=(0,\rho)^{\top}$, the initial magnetic
field
is the result of  $SU(2)\times U(1)$ vector-potential
remnants of the symmetric phase
 whose associated field tensor finds itself with a  non-zero projection along
the electromagnetic field after symmetry breaking.

There are several reasons to prefer the second gauge. One is that
in this gauge the constant
operator $Q=(\underline{\bf 1} + \tau^3)/2$
defines simple charge eigenstates
for all fields, while in the first gauge there is no
simple global definition of electric charge.
More importantly, the equivalence of gauges
holds only as long as $F^a_{\mu\nu}=0$.
As soon as the symmetry breaks, mass terms appear for the charged $W$ bosons
and for the $Z$ field, and the fields will start evolving into states with
non-zero
 $F^a_{\mu\nu}$.
Vector-field degrees of freedom of this type can no longer be transferred into
the Higgs field by a gauge transformation. Even in the
simplest case of a $U(1)$ symmetry, only the longitudinal degree of freedom of
the vector field can be exchanged with a phase of the Higgs field, while the
transverse
degrees of freedom are unaffected by gauge transformations.
The vector fields thus contain more dynamical
degrees of freedom than does the Higgs field. Therefore,
treating the issue of
generation of magnetic fields
from the point of view of the vector-boson fields is more appropriate.

Having shown that a magnetic field can be generated spontaneously
in the phase transition,
it remains to determine the initial strength and correlation length
of the field. Estimates of these two quantities are required in order
to predict the properties of the magnetic
seed field at the onset of galaxy formation.

To begin with, let us assume that the phase transition is
second-order with critical temperature $\tc$,
so that the Higgs expectation value for $T<\tc$ has the
generic temperature dependence $v(T)=v (1-T^2/\tc^2)^{1/2}$ where
$v=174.1$ GeV and $100$ GeV
$\lsim\tc\lsim 300$ GeV. The magnetic field will ``freeze out'', i.e.\
become insensitive to thermal fluctuations, at some temperature
$T_B<\tc$, where $T_B$ is to be determined in what follows.

Although the correlation length and strength of the magnetic field
can be calculated from the gauge-invariant
expression (\ref{Fdyn}), the computations
simplify considerably in the unitary gauge, where one may use
eq.~(\ref{Maxwell}) with ${\cal F}_{\mu\nu}^{\rm em}\equiv A_{\mu\nu}$.
Since the electromagnetic field $A_\mu$ is
massless, no natural scale emerges from the homogeneous part of the
equation,
$\partial^{\mu}A_{\mu\nu}=0$. The correlation length $\xi$ of the
magnetic field is instead determined by the source terms of
eq.~(\ref{Maxwell}). Because they
are at least quadratic in the charged fields $W_\mu$ and $W_\mu^{\dagger}$,
and since the
correlation length of each of
these fields is $M_W^{-1}(T_B)$ at the temperature $T_B$,
we find $\xi = [2 M_W(T_B)]^{-1}$.

Next, let us estimate the magnetic field strength. As the temperature
is lowered from $\tc$
to $T_B$, the Higgs potential energy density decreases by
the amount $\lambda [v(T_B)]^4= [v(T_B)]^2 [M_H(T_B)]^2/4$,
where $\lambda$ is the quartic Higgs coupling. The lost potential energy
is redistributed to the other positive definite terms
in the energy density. Let us assume that each such term receives
approximately the
same fraction $\gamma\sim 10^{-1}$ of the energy density.
In particular, we then have
\begin{equation}
\label{equipart}
|{\cal D}_{(\mu)}W_{(\nu)}-{\cal D}_{(\nu)}W_{(\mu)}|^2
\sim \frac{1}{2} g^2 v^2(T_B) |W_{(\mu)}|^2 \sim \gamma
\lambda [v(T_B)]^4~,
\end{equation}
where indices enclosed in parentheses indicate that there is no summation.
Here $W_\mu$ and ${\cal D}_\mu W_\nu$ are defined below
eq.~(\ref{MaxwellUG}).
Inserting these estimates and the expression for $\xi$ into equation
(\ref{Maxwell}) we obtain
\begin{equation}
\label{Bspont}
B\sim \frac{\sintw}{g} \gamma M_H^2(T_B)~.
\end{equation}

In Ref.~\cite{Martin} it was argued that magnetic fields
become stable to thermal fluctuations when the temperature drops
below the so-called Ginzburg temperature $T_{\rm G}$,
defined as the temperature below which thermal fluctuations
become too weak to restore the symmetry locally.
Here we shall provide a more conservative estimate, and assume that
the magnetic field
freezes out at a temperature $T_B<T_{\rm G}$ when the
typical energy of thermal fluctuations
drops below the
magnetic field energy contained within a correlated
domain of approximate volume $\xi^3$.
Hence, $T_B$ is determined by the condition $\xi^3 B^2/2\sim T_B$.
Inserting the characteristic temperature dependence of the masses,
we obtain for a second-order phase transition
\begin{equation}
\label{tb2}
\frac{T_B^2}{\tc^2} \sim
\left[ 1 + \frac{\tc^2}{M_W^2}\left( \frac{4 g}{\gamma\sintw}
\frac{M_W^2}{M_H^2}\right)^{\!4}\,\right]^{-1}~.
\end{equation}
Thus, except in the case of a very high value of the Higgs-boson mass,
$T_B/\tc$ is not near unity, and masses at temperature $T_B$
 are well approximated by their values at zero temperature.

If the phase transition is first-order, magnetic fields can still be
generated spontaneously within the bubbles of broken phase.
The masses for $T<\tc$
are close to their values at zero temperature. From the condition
$\xi^3 B^2/2\sim T_B$ we obtain for a first-order phase transition
\begin{equation}
\label{tb1}
T_B\sim\left(\frac{\gamma\sintw}{4g}\frac{M_H^2}{M_W^2}
\right)^{\!2} M_W~.
\end{equation}
This temperature approximates that of eq.~(\ref{tb2})
in the limit of a low Higgs-boson mass.
Note
that the magnetic
freeze-out temperature $T_B$ is generally lower than the
Ginzburg temperature, which may be very close to the critical
temperature \cite{Grasso}.
This is to be expected, since the
magnetic field contains only part of the energy released by the
Higgs potential in the phase transition and thus may be destroyed by
smaller thermal fluctuations.

In summary, for spontaneously generated magnetic fields in either
a first- or second-order electroweak phase transition,
the above estimates of the magnetic field strength $B$ and
correlation length $\xi$ give
\begin{equation}
\label{spontest}
B_{\rm sp.} \sim \left[\frac{M_H}{100\mbox{~GeV}}\right]^2\cdot
10^{22}~{\rm Gauss}~,\quad\quad \xi_{\rm sp.}\sim 10^{-2} {\rm GeV}^{-1}~.
\end{equation}
These estimates are in rough agreement with those derived in
Ref.~\cite{Vachaspati}.

\section{Magnetic Fields from Bubble Collisions}
\label{bubbles}
Let us now consider the possibility of forming magnetic fields in the
collisions
of two bubbles of broken vacuum
in a first-order electroweak phase transition. Such collisions
were
investigated in Refs.~\cite{Copeland2,Grasso}
for some special cases. Using the
same model as those references for the initial evolution,
I shall show here that no magnetic field is
generated for arbitrary difference and relative orientation of the Higgs
phases of the two bubbles.

For initial times, the Higgs field configurations of  two disjoint bubbles of
arbitrary shape and size are, respectively,
\begin{equation}\label{initbub}
\Phi^1_i(\bm{x}) = \left( \begin{array}{c}0\\ \rho_1(\bm{x}) \end{array}
\right)
\quad {\rm and}\quad
\Phi^2_i(\bm{x}) = \exp\left[i\frac{\theta_0}{2} n^a\tau^a \right]
\left( \begin{array}{c}0\\ \rho_2(\bm{x}) \end{array} \right)~,
\end{equation}
where $\hat{\bm{n}}=(n^1,n^2,n^3)$ is a constant unit vector.
The phases and orientations of the Higgs field within each bubble have
equilibrated to constant values. A constant $U(1)_Y$ factor $e^{i\varphi_0}$
was excluded
from $\Phi^2_i$, since $\varphi_0$ can be absorbed into $\theta_0 n^3/2$.
Because $n^a\tau^a$ is the only
Lie-algebra direction  involved,
one may write the initial complete Higgs field
as \cite{Copeland2}
\begin{equation}\label{inittot}
\Phi_i(\bm{x}) = \exp\left[i\frac{\theta(\bm{x})}{2} n^a\tau^a\right]
\left( \begin{array}{c}0\\ \rho(\bm{x})\\ \end{array} \right)~.
\end{equation}

Furthermore, the authors of \cite{Copeland2,Grasso}
have assumed that all gauge potentials
and their derivatives vanish initially. As we learned in the preceding section,
one is free to choose such a gauge as long as
 the field tensors $F^a_{\mu\nu}$
and  $F^Y_{\mu\nu}$ also vanish.

Proceeding as the references, we assume that the above expressions are valid
until
the two bubbles collide. One may easily evaluate $\hat{\phi}^a$ which may
be written as $\bm{\hat{\phi}} = \cos\theta\, \bm{\hat{\phi}}_0 +
\sin\theta\,
\hat{\bm{n}}\times \bm{\hat{\phi}}_0 + 2 \sin^2\!\frac{\theta}{2}\,
 (\hat{\bm{n}}\cdot \bm{\hat{\phi}}_0) \hat{\bm{n}}$
where $\bm{\hat{\phi}}_0 = (0,0,-1)^{\top}$.
Then
$\partial_\mu\hat{\phi}^a$ takes the particularly simple form
$\partial_\mu\bm{\hat{\phi}} = \partial_\mu \theta\,
\hat{\bm{n}}\times \bm{\hat{\phi}}$.  The
last
term of the electromagnetic field ${\cal F}^{\rm em}_{\mu\nu}$
(eq.~(\ref{goodone}))
thus vanishes, and since
$F^a_{\mu\nu}$ and  $F^Y_{\mu\nu}$ are zero, the electromagnetic field
vanishes.
Similarly, the electric current (\ref{Fdyn}) vanishes.

It is instructive to check this result by transforming the Higgs field into the
unitary gauge,
using the group-valued function $\Omega = U= \exp[-i n^a\tau^a \theta/2]$.
This leads to a vector potential of
the
form (\ref{symmvac}).
It  follows easily from eqs.~(\ref{trick}), or alternatively from
eqs.~(\ref{uopar}) and (\ref{magspont}),
that  their contribution to the magnetic field is zero.  From the latter of
these equations
it is apparent that the phases of our $U$ are rather special, and
that there in general  would be a magnetic field. The absence of a magnetic
field
can be traced directly to the
fact that the unit vector $\hat{\bm{n}}$ is a constant or, more precisely,
that the Higgs phases depend on only one parameter $\theta$.
In Ref.~\cite{Copeland2} it was proven that the Higgs field in
any two-bubble collision can  be written in the form (\ref{inittot}) for
constant $n^a$.
We thus conclude that no magnetic field is generated from the initial
classical evolution of the Higgs field in an electroweak two-bubble
collision. One should remember, though, that the expression
(\ref{inittot})  is probably too simplistic to describe what takes place
once the bubbles overlap significantly. Magnetic fields could then
emerge gradually if $\hat{\bm{n}}$ develops a spatial dependence.

The present result is in stark contrast to that of the abelian $U(1)$ model
\cite{KibVil,Ahonen},
in which a field strength is present from the instant of collision.
The principal difference is that the
$U(1)$ vector field in that model is massive and the corresponding field
strength is generated  as a result of the coupling of
the $U(1)$ field  to the Higgs field. In contrast, the electromagnetic
$U(1)$ field in the broken electroweak theory is distinguished as that
direction of the
Lie algebra that does {\em not\/} couple to the Higgs field.

In the electroweak theory, in order to generate a magnetic field already
at the instant of bubble collision,
one would need an initial configuration in which $\hat{\bm{n}}$
has a spatial
dependence, i.e.\ where the Higgs phases are generated by at
least two elements of the Lie algebra. The simplest example of this
would be a three-bubble collision \cite{Ambleside}.

If one relaxes the assumption that the gauge potentials are zero initially,
magnetic fields may also emerge spontaneously within each bubble by the
mechanism described in the previous section.
When the presence of the plasma is taken into account, other processes may
lead to the creation of magnetic fields. In particular, magnetic fields may
stem
from the motion of dipole charge layers that
develop on bubble walls because of the baryon asymmetry \cite{Baym}.
It is  also possible that bubble collisions give rise to field
configurations
which indirectly produce magnetic fields. This will be investigated in
Section \ref{topdef}.

Let us now provide some coarse
estimates of the strength
and correlation length of the magnetic field, assuming only that
it arises from
some mechanism directly associated with bubble collisions.

The growth of nucleated bubbles in the electroweak phase transition
has
recently been studied numerically by Kurki-Suonio and Laine \cite{Kurki},
using model parameters obtained from lattice computations.
For a weak first-order transition with $M_H>68~{\rm GeV}$, they find that
the average radius of the bubbles at the time of collision is
$R\lsim 10^{-7} t_{\rm EW}\sim 10^6~{\rm GeV}^{-1}$, where
$t_{\rm EW}$ is the Horizon scale at the time of the phase transition.
Under the assumption that the magnetic field is coherent on the scale
of a bubble radius, we may take $\xi\sim R$ as the correlation length.
One can then derive a naive estimate of the magnetic field strength,
using $\oint A_i\,dx_i$ as the expression for the magnetic flux
enclosed by a loop the size of a bubble. Noting that $A_i\sim \partial_i
\vartheta/g$ for some angle $\vartheta$ on the
Higgs vacuum manifold $S^3$,
and that the average difference in the value of
$\vartheta$ between two adjacent bubbles is of the order of $\pi$,
we obtain $BR^2\sim \pi/g$, and thus a lower bound
on the magnetic field, $B\gsim 10^8$ Gauss. The correlation length
of this field is much larger, and
the strength much smaller,
than for fields arising from
other mechanisms described in this article. Nevertheless,
the results of
Ref.~\cite{Kurki} indicate that an even weaker first-order
phase transition leads to smaller bubbles, and hence a larger
magnetic field.

  On the other hand, taking into account the effects of the finite
conductivity of the
plasma after two of the bubbles have initially touched,
Ahonen and Enqvist \cite{Ahonen}
argue, subject to some approximations, that diffusion causes
magnetic flux to become concentrated around the expanding
circle of most recent intersection of the two bubbles. In this way,
they predict a correlation length of $\xi\sim 10^4~{\rm GeV}^{-1}$
and a magnetic field strength of $B\sim~10^{20}$ Gauss.

We see that the presence of the plasma may have a dramatic effect on
the order of magnitude of the magnetic field.
A numerical field-theory simulation of multi-bubble collisions in the full
electroweak theory
is currently in progress, initially neglecting the plasma.
Once the results of this simulation are known and understood,
the various plasma effects can be incorporated.
It should then become possible to provide more precise
predictions of the strength and correlation length of magnetic fields
that arise in electroweak bubble collisions.

\section{Magnetic Fields from Non-Topological Defects}
\label{topdef}
It was recently suggested by Grasso and Riotto \cite{Grasso} that
magnetic fields may arise from
non-topological defects formed in the electroweak phase transition,
such as Z-strings \cite{Zref} and W-strings \cite{embdef}.  These are
string-like
embedded
vortex solutions of the electroweak theory characterized by the
winding of a phase of the Higgs field around a core where the Higgs field
goes to zero.  The core encloses a flux quantum of one of the gauge-field
components which attains considerable field strength, since the characteristic
width is given by the inverse vector-boson mass. In a $U(1)$ model, these
defects
are
topologically stable, but in the electroweak theory the phase
can unwind by slipping over the simply connected vacuum manifold, and the
defect decays to the vacuum.

Saffin and Copeland  \cite{Copeland2} have shown that
$W$-string and $Z$-string configurations may be
generated during bubble collisions in the $SU(2)_{\rm L}\times U(1)_{Y}$
theory.
In terms of the notation of the previous section, this occurs in the two
special cases when
the unit vector $\hat{\bm{n}}$
is perpendicular or parallel to $\bm{\hat{\phi}}_0$,
respectively.
In these cases, the effective symmetry group of the problem reduces to $U(1)$,
for which vortex production in bubble collisions has been studied earlier
\cite{KibVil,Ahonen}.
In simulations the strings form as circular loops along the circle
of intersection of the two bubbles, with the axis of the loop coinciding with
the
line through the two bubble centers.

There are three important questions that need be answered in connection
with the possible generation of magnetic fields from non-topological defects.
\begin{itemize}\itemsep=0pt\parsep=0pt\parskip=0pt
\item[1.] Do the defects themselves carry magnetic fields?
\item[2.] Do the defects contain electrically charged fields which could
produce
 electric currents?
\item[3.] Are electromagnetic fields generated when these unstable
defects decay?
\end{itemize}
I shall defer the last question to the end of this section and  begin  instead
to
address
the first two questions. For a reasonable set of
definitions, and in the absence of magnetic monopoles, they should be
equivalent.

In defiance of  such expectations,  some surprising results were recently
obtained in  Ref.~\cite{Grasso}. The results seemed to indicate
that a magnetic field would always be present along the internal axis of a
$Z$-string, which is known to contain only neutral fields.
This interpretation was based on the conventional gauge-invariant
definition of the electromagnetic field
tensor, eq.~(\ref{HooftSM}), which led to the inclusion of the last term of
eqs.~(\ref{badf}),  (\ref{MaxwellUG}) in the unitary gauge.

As we have learned in section II, there exist alternative
definitions of
the electromagnetic field tensor which coincide only
when the magnitude of the Higgs field
is constant. I have argued that the definitions of the field tensor and
electric current
given in eqs.~(\ref{goodone}) and (\ref{Fdyn}) are more appropriate,
in that  ${\cal F}^{\rm em}_{\mu\nu}$ always reduces to $A_{\mu\nu}$ in the
unitary
gauge and electrically neutral fields never serve as sources for the
electromagnetic field. Indeed,
with the new definitions  everything becomes perfectly consistent
with naive expectations. In order to illustrate this,
let us  investigate the  field configurations for
the $Z$- and $W$-strings in some detail. They can be written in the form
\begin{equation}\label{Zdef}
\underline{A}^Z_{\,\varphi}=  \frac{mv(r)}{r} \left(\begin{array}{cc}\cos
2\tw&0\\
0&-1 \end{array}\right)~,
\quad \Phi^Z = \rho(r)
\left(\begin{array}{c}0\\
e^{im\varphi} \end{array}\right)~,
\end{equation}
and
\begin{equation}\label{Wdef}
\underline{A}^W_{\,\varphi}=  \frac{m\tilde{v}(r)}{r}
\left(\begin{array}{cc}0&e^{i\delta}\\
e^{-i\delta}&0 \end{array}\right)~,
\quad \Phi^W = \tilde{\rho}(r)
\left(\begin{array}{c}i e^{i\delta} \sin{m\varphi}\\
\cos m\varphi \end{array}\right)~,
\end{equation}
where $r,\varphi$ are cylindrical coordinates, $\delta$ is an arbitrary real
number labeling
a family of gauge-equivalent $W$ vortex solutions, and
$m$ is the integer winding number. Because of its particular phase singularity
at $r=0$,
there is no non-singular expression for the $W$ vortex in a gauge
where the upper component of the Higgs field is zero
\cite{NOstabil}.

For the $Z$-string configuration, we obtain $\hat{\phi}^a=-\delta^{a3}$, and
thus the
last term of eq.~(\ref{goodone}) vanishes. The first two terms combine to give
$\sintw \partial_{[\mu} W^3_{\nu]} + \costw F^Y_{\mu\nu} = 0$  and so ${\cal
F}^{\rm em}_{\mu\nu}$
vanishes. With the electric
current, eq.~(\ref{Fdyn}), we find that
$(D^\mu \hat{\phi})^3 =F^1_{\mu\nu}=F^2_{\mu\nu}=0$, and the last term is just
a derivative of the term we previously found to be  zero, so there is no
electric current.

Next, let us investigate the $W$-string solution. It is convenient to recognize
that
it is of the form  $\underline{A}_\varphi=m  n^a\tau^a \tilde{v}(r)/r $ and
$\Phi=
\exp[im\varphi n^a\tau^a] (0,\tilde{\rho}(r))^{\top}$ for
$\hat{\bm{n}}=(\cos\delta,
-\sin\delta,0)$.
Using the method of the
previous section, we find $\bm{\hat{\phi}} =
\cos (2 m\varphi)\, \bm{\hat{\phi}}_0 + \sin (2 m\varphi)\,
\hat{\bm{n}}\times \bm{\hat{\phi}}_0 + 2 \sin^2\!(m\varphi)\,
 (\hat{\bm{n}}\cdot \bm{\hat{\phi}}_0) \hat{\bm{n}}$
 where $\bm{\hat{\phi}}_0 = (0,0,-1)^{\top}$.
The only non-zero field-tensor
components are $F^a_{r\varphi}= [m \tilde{v}'(r)/r] n^a$. Because
$n^a\hat{\phi}^a \equiv n^a\hat{\phi}_0^a = 0$, we have that the term
$\hat{\phi}^a F^a_{r\varphi} = 0$ in  eq.~(\ref{goodone})  vanishes. In the
last term of this equation,
one of the  factors is $\partial\hat{\phi}^b/\partial r = 0$. Thus
${\cal F}^{\rm em}_{r\varphi}$ vanishes.

The issue of whether there is an electric current is more interesting in
the
case of the $W$-string, since its gauge fields involve  charged fields
$W^1_\varphi$ and
$W^2_\varphi$. On the other hand, also the phases of the Higgs field are
charged,
as compared with the unitary-gauge vacuum.  We find the last term of the
current
(\ref{Fdyn}) to be zero as before.  Since
$\partial\hat{\phi}^a /\partial r =
0$ and
there is no radial component $\underline{A}_r$, only the
$r$-component
of the current may be non-vanishing. We now make use of the relation
$\partial_\varphi
\bm{\hat{\phi}} = 2 m \hat{\bm{n}}\times\bm{\hat{\phi}}$ and can write
$(D_\varphi\bm{\hat{\phi}}) =
2 m [(1 + v(r))/r]\, \hat{\bm{n}}\times\bm{\hat{\phi}}$.
This is perpendicular to
$\hat{\bm{n}}$,
and so the  term $(D_\varphi\hat{\phi})^a F^a_{\varphi r}$ vanishes, and there
is no electric current.

Although this section has so far only confirmed what was expected,
it has served as a nice illustration of the
 properties and applicability  of the new definition of the electromagnetic
field tensor
${\cal F}^{\rm em}_{\mu\nu}$. We have established that it works and
that it gives results that are reasonable in cases where the conventional
definition appears to lead to absurdities.

Finally, I shall discuss the suggestion made in Ref.~\cite{Grasso}
that magnetic
fields may be generated in the decay of  $Z$-strings.
It is well-known that the
unstable $Z$-string decays initially through charged $W$-boson fields
\cite{Perkins,NOstabil}.
The idea is that these $W$ fields form a ``condensate'' which
then in turn would act as a source of magnetic fields.
One extremely  important caveat  is that  the presence of $W$ fields is
highly transient, as
the $Z$-string is known to
decay to a vacuum configuration \cite{Achucarro}.
It is conceivable, however, that the large conductivity
of the plasma in the early universe \cite{ZelPark,TurWid,Baymetal}
may cause the magnetic field lines to freeze into the fluid so
that it remains preserved at later times.

The instability of the $Z$-string is a result of the occurence in
the energy density of a term
\begin{equation}\label{coup}
i g \costw Z_{12}   (W_1^{\dag} W_2 - W_2^{\dag} W_1)
\end{equation}
which couples the
field strength $Z_{12} =\partial_{1} Z_{2} -\partial_{2}
Z_{1} $
with the magnetic dipole moment of the $W$ boson.
 The energy is lowered through a suitable alignment of
this magnetic moment,
corresponding to $W_1=-iW_2 \equiv W$ for $Z_{12}>0$.
The instability is greatest at the
center
of the vortex, where $Z_{12}$ is largest and where the $W$ mass term is reduced
by the
vanishing of the Higgs field.  Let us make the simplified assumption that
$Z_{12}$ is approximately uniform in the core of the vortex. This is actually
justified if the Higgs-boson mass is considerably larger than the $Z$-boson
mass.
In such a case, the unstable modes of the $W$ field are well-known
\cite{Ambjorn,MacDowell}. The mode that peaks in the center of the
vortex is
given by
\begin{equation}\label{Wcond}
W(r) = W(0) \exp (-\frac{1}{4} g C r^2)~,
\end{equation}
where $C=\costw Z_{12}$. For this mode, it is easy to check that
$F^1_{ij}=F^2_{ij}= 0$. This is in fact true for any unstable mode
\cite{Ambjorn,MacDowell}.
Neglecting back reactions on the Higgs field, we
still have $\hat{\phi}^a=-\delta^{a3}$.
The last term of eq.~(\ref{goodone}) evaluates to $2e|W|^2$ which cancels
against parts of the first term, leaving
 ${\cal F}^{\rm em}_{ij}=A_{ij}$ as usual.
In the current eq.~(\ref{Fdyn}) something more interesting happens.
Since $(D_i \hat{\phi})^3 = 0$, we are left only with the last term, and the
equation for the magnetic field can be written
\begin{equation}\label{magnet}
\partial_i ~({\cal F}^{\rm em}_{12} - 2e|W|^2) = 0~.
\end{equation}
The (non-uniform) magnetic field $B={\cal F}^{\rm em}_{12}$
is thus entirely comprised of the magnetization from the
$W$ bosons. It is apparent that the $W$ bosons initially present in the
decay of the $Z$-string do indeed generate a magnetic field.

Let us now compute the strength and correlation length
of the magnetic field produced by this mechanism. An upper bound on
the magnitude of
$|W|^2$ can be obtained by studying the growth of the $W$ field
in a fixed Z-string background, which is limited by the
quartic term in the energy density, $g^2 |W_1^{\dagger} W_2-
W_2^{\dagger} W_1|^2/2$. The instability ceases at a maximal value
$|W|^2 = \costw Z_{12}/g$. From eq.~(\ref{magnet}) one then obtains the
bound
\begin{equation}
\label{BfromZ}
B\lsim \sin 2\tw Z_{12}~.
\end{equation}
To find an estimate for $Z_{12}$ one can use the flux quantization
condition $\int d^2\!x Z_{12} = 4\pi \costw/g$,
which arises from requiring the
covariant derivative of the Higgs field to vanish asymptotically.
The integral here is evaluated over a surface perpendicular to the
$Z$-string.
Assuming that the flux
is confined to an approximate cross-sectional area $\pi M_Z^{-2}$,
we find $Z_{12}\sim 4\costw M_Z^2/g$, and therefore
\begin{equation}
\label{Fexpl}
B\lsim \frac{8\cos^2\tw \sin\tw}{g} M_Z^2~.
\end{equation}
The growth of $W$ fields, and therefore of a correlated magnetic
field, is limited to the region where $Z_{12}$ is
large and $\Phi^{\dagger}\Phi$ is small. These regions have
characteristic widths $M_Z^{-1}$ and $M_H^{-1}$, respectively.
The correlation
length of the magnetic field is thus $\xi\sim\min(M_Z^{-1}, M_H^{-1})
\sim M_H^{-1}$, since experimentally $M_H>77.5$ GeV (95 \% C.L.)
\cite{Sopczak}.

In summary, for magnetic fields generated by decaying non-topological
defects, we obtain the following numerical estimates:
\begin{equation}
\label{nontopest}
B_{\rm ntop.} \lsim
10^{24}~{\rm Gauss}~,\quad\quad \xi_{\rm ntop.}\sim
\left[\frac{100~{\rm GeV}}{M_H}\right] \cdot 10^{-2}~{\rm GeV}^{-1}~.
\end{equation}

\section{Conclusions}

The main results of this paper are as follows:  I have
established that magnetic fields are  indeed  generated classically from
Higgs and gauge fields in the electroweak
phase transition through the mere process of spontaneous symmetry breaking,
as was originally suggested by Vachaspati \cite{Vachaspati}.
Reformulating the problem in the unitary gauge, I have explicitly constructed
the magnetic field thus generated.  Previous claims that no such magnetic field
is
produced  were based on an investigation in which
an incomplete expression for the electric current from the Higgs
field was used,
and currents from charged vector bosons were neglected.

Moreover, I have shown that no magnetic field results initially from the
classical
evolution of the Higgs field in a collision of  two bubbles in a first-order
electroweak phase transition. This was shown for arbitrary difference
and relative orientation of the phases of the Higgs field.  The reason is
that only one constant direction in the Lie algebra is involved.
Nevertheless, one should not exclude the possibility that later evolution
of the fields could give rise to magnetic fields. These issues are currently
being
investigated.

Furthermore, I have pointed out that the notion of an electromagnetic field
tensor is ambiguous whenever the magnitude of the Higgs field is not constant.
With the conventional gauge-invariant
definition, eq.~(\ref{HooftSM}),  electrically neutral currents
may give rise to electromagnetic fields. In particular,
magnetic fields may be present inside electrically neutral configurations such
as the
$Z$-string. In order to remedy this, I have proposed a
different gauge-invariant
definition of the electromagnetic field, eq.~(\ref{goodone}), which ensures
that
no electromagnetic fields are generated from neutral sources and
which coincides with
the other definition for constant Higgs-field magnitude.

The issue of the definition of the electromagnetic field tensor is important
for
the interpretation and description of physical phenomena, but should have
no bearing on the physics, as the various fields evolve independently of how we
interpret them. One particular example concerns the simultaneous collision of
multiple similar-sized bubbles at the time of percolation,
after which the Higgs magnitude is expected to fluctuate violently
\cite{Rocky}.
In the presence of $Z$ fields one would then conclude from
eq.~(\ref{MaxwellUG}),
which follows from definition (\ref{HooftSM}),
that electromagnetic fields are created from the gradients of this magnitude.
In such a context it is important to realize that
any statement about the presence or absence
of electromagnetic fields will depend
on which definition of the electromagnetic field tensor is used,
and agreement will only be reached if
the evolution of all fields is traced
to a later time when the Higgs magnitude has assumed
a uniform value. Nevertheless, if one makes the assumption that
the Higgs field relaxes to a uniform value without exciting any new dynamics
in the angular degrees of freedom, the new definition (\ref{goodone}) has the
property that it predicts
the same magnetic field during the fluctuating stage as it does after the
fluctuations of
the Higgs magnitude have
subsided.

Finally, I have verified that a magnetic field is produced in the initial decay
of the $Z$ string, as was suggested in Ref.~\cite{Grasso}. Although
such a field  is transient in the pure Yang-Mills-Higgs model, it is
conceivable that
it may survive until later times due to the high conductivity of the plasma in
the
early universe.

Estimates of the strength and correlation length of the initial
magnetic field
have been provided for each of the three mechanisms of production:
Spontaneous generation, bubble collisions, and decay of non-topological
defects. The subsequent evolution of correlated domains may
be calculated according to the recipe presented in
Ref.~\cite{DimDav}. The correlation length may increase faster than
the scale factor $a(t)$
due to the presence of magnetohydromagnetic
Alfv\'{e}n waves \cite{DimDav}. Such waves serve to bring
two initially uncorrelated
domains into causal contact, so that the magnetic field
may untangle and smoothen.

With the possible exception of bubble collisions, one finds in all
cases that the correlation length at the time of equal matter and
radiation energy densities, $t_{\rm eq.}\sim 10^{11}$ sec,
is smaller than the magnetic diffusion length
$l_{\rm d}\sim 10^{23}$ GeV$^{-1}$. This remains true also
when enhancement due to Alfv\'{e}n waves is taken into account.
In order to evolve into a seed field of sufficient correlation length and
strength at the onset of galaxy formation,
fields of such weak correlation may require, depending on how
the root-of-mean-square average of the magnetic field is
calculated \cite{Hind2},
some additional mechanism which stretches the correlation length,
such as non-linear inverse cascade
\cite{Brandenburg}.

In the case of bubble collisions it is still an open question
whether one may
produce
a correlated, strong magnetic field
without the need to invoke complicated models of magnetohydrodynamic
turbulence such as non-linear inverse cascade.
This issue is likely to be
resolved with the results from current and future
computer simulations of bubble collisions in the electroweak theory.

\subsection*{ACKNOWLEDGMENTS}
I am indebted to
T.~Vachaspati, A.-C. Davis,
D.J.H.~Chung, E.J.~Copeland,
M.~Hindmarsh, N.~Manton, A.~\mbox{Riotto,}
and P.M.~Saffin for constructive discussions and valuable comments.
The original work was supported in part by
the Swedish Natural Science Research Council (NFR) and in part by
DOE and NASA
under Grant NAG5--2788. Additional results included in the
final version of the manuscript were obtained with support from
EPSRC under Grant GR/K50641.

\def\apj#1#2#3{{\it Astrophys.\ J.\ }{{\bf #1}, {#3}\ {(#2)}}}
\def\app#1#2#3{{\it Astropart.\ Phys.\ }{{\bf #1}, {#3}\ {(#2)}}}
\def\np#1#2#3{{\it  Nucl.\ Phys.\ }{{\bf #1}, {#3}\ {(#2)}}}
\def\pr#1#2#3{{\it Phys.\ Rev.\ }{{\bf #1}, {#3}\ {(#2)}}}
\def\pl#1#2#3{{\it  Phys.\ Lett.\ }{{\bf #1}, {#3}\ {(#2)}}}
\def\prl#1#2#3{{\it Phys.\ Rev.\ Lett.\ }{{\bf #1}, {#3}\ {(#2)}}}
\def\prep#1#2#3{{\it Phys.\ Rep.\ }{{\bf #1}, {#3}\ {(#2)}}}
\def\jp#1#2#3{{\it J.\ Phys.\ A\ }{{\bf #1}, {#3}\ {(#2)}}}
\def\ibid#1#2#3{{\it ibid.\  }{{\bf #1}, {#3}\ {(#2)}}}
\def\mpl#1#2#3{{\it Mod.\ Phys.\ Lett.\ }{{\bf #1}, {#3}\ {(#2)}}}

\end{document}